\documentclass[12pt,preprint]{aastex}

\shorttitle{}
\shortauthors{Levi, Kenyon, Podolak and Prialnik}

\usepackage{graphicx}
\usepackage{textcomp}
\usepackage{amssymb}
\usepackage{natbib}
\usepackage{amsmath}
\usepackage{subfigure}
\usepackage{multirow}\usepackage[section] {placeins}
\def\deg{\ifmmode {^\circ}\else {$^\circ$}\fi}
\def\degree{\ifmmode {^\circ}\else {$^\circ$}\fi}
\def\mum{\ifmmode {\rm \,\mu {\rm m}}\else $\rm \,\mu {\rm m}$\fi}
\def\arcsec{\ifmmode ^{\prime \prime}\else $^{\prime \prime}$\fi}

\def\inch{\ifmmode ^{\prime \prime}\else $^{\prime \prime}$\fi}
\def\msunyr{\ifmmode {M_{\odot}~{\rm yr^{-1}}}\else $M_{\odot}~{\rm yr^{-1}}$\fi}
\def\msun{\ifmmode {M_{\odot}}\else $M_{\odot}$\fi}
\def\rsun{\ifmmode {R_{\odot}}\else $R_{\odot}$\fi}
\def\lsun{\ifmmode {L_{\odot}}\else $L_{\odot}$\fi}
\def\mstar{\ifmmode {M_{\star}}\else $M_{\star}$\fi}
\def\rstar{\ifmmode {R_{\star}}\else $R_{\star}$\fi}
\def\tstar{\ifmmode {T_{\star}}\else $T_{\star}$\fi}
\def\lstar{\ifmmode {L_{\star}}\else $L_{\star}$\fi}
\def\md{\ifmmode {M_d}\else $M_d$\fi}
\def\ld{\ifmmode {L_d}\else $L_d$\fi}
\def\ad{\ifmmode A_d\else $A_d$\fi}
\def\ldlstar{\ifmmode L_d / L_\star\else $L_d / L_{\star}$\fi}
\def\rearth{\ifmmode {\rm R_{\oplus}}\else $\rm R_{\oplus}$\fi}
\def\mearth{\ifmmode {\rm M_{\oplus}}\else $\rm M_{\oplus}$\fi}
\def\qdstar{\ifmmode Q_D^\star\else $Q_D^\star$\fi}
\def\kms{\ifmmode {\rm km~s^{-1}}\else $\rm km~s^{-1}$\fi}
\def\ms{\ifmmode {\rm m~s^{-1}}\else $\rm m~s^{-1}$\fi}
\def\mesc{\ifmmode m_{esc}\else $m_{esc}$\fi}
\def\rmin{\ifmmode r_{min}\else $r_{min}$\fi}
\def\rmax{\ifmmode r_{max}\else $r_{max}$\fi}
\def\mmin{\ifmmode m_{min}\else $m_{min}$\fi}
\def\mmax{\ifmmode m_{max}\else $m_{max}$\fi}
\def\rmind{\ifmmode r_{min,d}\else $r_{min,d}$\fi}
\def\rmaxd{\ifmmode r_{max,d}\else $r_{max,d}$\fi}
\def\mmaxd{\ifmmode m_{max,d}\else $m_{max,d}$\fi}
\def\vrad{\ifmmode v_{rad}\else $v_{rad}$\fi}
\def\qz{\ifmmode q_{0}\else $q_{0}$\fi}
\def\qi{\ifmmode q_{i}\else $q_{i}$\fi}
\def\ql{\ifmmode q_{l}\else $q_{l}$\fi}
\def\qs{\ifmmode q_{s}\else $q_{s}$\fi}
\def\rbrk{\ifmmode r_{brk}\else $r_{brk}$\fi}
\def\rdamp{\ifmmode r_{damp}\else $r_{damp}$\fi}
\def\rin{\ifmmode r_{in}\else $r_{in}$\fi}
\def\rout{\ifmmode r_{out}\else $r_{out}$\fi}
\def\tin{\ifmmode t_{in}\else $t_{in}$\fi}
\def\tout{\ifmmode t_{out}\else $t_{out}$\fi}
\def\ain{\ifmmode a_{in}\else $a_{in}$\fi}
\def\aout{\ifmmode a_{out}\else $a_{out}$\fi}
\def\r0{\ifmmode R_{0}\else $R_{0}$\fi}
\def\m0{\ifmmode m_{0}\else $m_{0}$\fi}
\def\M0{\ifmmode M_{0}\else $M_{0}$\fi}
\def\xm{\ifmmode x_{m}\else $x_{m}$\fi}
\def\sigz{\ifmmode \Sigma_0\else $\Sigma_0$\fi}
\def\gyr{\ifmmode {\rm g~yr^{-1}}\else ${\rm g~yr^{-1}}$\fi}
\def\cms{\ifmmode {\rm cm~s^{-1}}\else ${\rm cm~s^{-1}}$\fi}
\def\gcms{\ifmmode {\rm g~cm^{-2}}\else $\rm g~cm^{-2}$\fi}
\def\gcmc{\ifmmode {\rm g~cm^{-3}}\else $\rm g~cm^{-3}$\fi}
\def\ecsk{\ifmmode {\rm erg~cm^{-1}~s^{-1}~K^{-1}}\else $\rm erg~cm^{-1}~s^{-1}~K^{-1}$\fi}
\def\atilin{\ifmmode {\tilde{a}_{in}}\else $\tilde{a}_{in}$\fi}
\def\atilout{\ifmmode {\tilde{a}_{out}}\else $\tilde{a}_{out}$\fi}
\def\atil{\ifmmode {\tilde{a}}\else $\tilde{a}$\fi}
\def\ttil{\ifmmode {\tilde{t}}\else $\tilde{t}$\fi}
\def\sqrttt{\ifmmode {\tilde{t}^{1/2}}\else $\tilde{t}^{1/2}$\fi}

\begin{document}
\pagenumbering{arabic}

\title{H-Atmospheres of Icy Super-Earths Formed in situ in the Outer Solar System: An Application to a Possible Planet Nine}
\author{A. Levi\altaffilmark{1}, S.J. Kenyon\altaffilmark{1}, M. Podolak\altaffilmark{2}, and D. Prialnik\altaffilmark{2}}
\altaffiltext{1}{Harvard-Smithsonian Center for Astrophysics, 60 Garden Street, Cambridge, MA 02138, USA}
\altaffiltext{2}{Department of Geosciences, Tel Aviv University, Tel Aviv, Israel 69978}
\email{amitlevi.planetphys@gmail.com}

\maketitle

\section*{ABSTRACT}

We examine the possibility that icy super-Earth mass planets, formed over long time scales (0.1--1~Gyr) 
at large distances ($\sim$ 200--1000~AU) from their host stars, will develop massive H-rich atmospheres. 
Within the interior of these planets, high pressure converts CH$_4$ into ethane, butane, or diamond and 
releases H$_2$. Using simplified models which capture the basic physics of the internal structure, we
show that the physical properties of the atmosphere depend on the outflux of H$_2$ from the mantle.  When 
this outflux is $\lesssim 10^{10}$\,molec\,cm$^{-2}$\,s$^{-1}$, the outgassed atmosphere has base pressure
$\lesssim$ 1 bar. Larger outflows result in a substantial atmosphere where the base pressure may approach 
$10^3 - 10^4$ bar. For any pressure, the mean density of these planets, 2.4--3~g~cm$^{-3}$, is much larger
than the mean density of Uranus and Neptune, 1.3--1.6~g~cm$^{-3}$. Thus, observations can distinguish 
between a Planet Nine with a primordial H/He-rich atmosphere accreted from the protosolar nebula and one 
with an atmosphere outgassed from the core.
 
%\bibliographystyle{apj}

%The hypothesized Planet Nine may be a scattered mini-Neptune or it may have slowly formed {\it in situ} as a water rich super-Earth. In the latter mode of formation it probably missed the opportunity to capture H/He from the nebula. However, high-pressure processing of hydrocarbons into diamond within the water rich mantle releases hydrogen that may outgas, yielding a \textit{non-primordial} hydrogen atmosphere. We find that for water rich planets more massive than $\approx 10~\mearth$ internally produced hydrogen, if allowed to completely outgas, may result in a secondary hydrogen atmosphere as extensive as of the order of $10^5$\,bar. However, we find that the outgassing of hydrogen is restricted. The formation of clathrate hydrates of hydrogen at relatively shallow depths can entrap atmospheric hydrogen, yielding a sub-bar atmosphere only. This mechanism is dominant if the outflux of internal hydrogen is on average less than $10^{10}$\,cm$^{-2}$\,s$^{-1}$. For a larger outflux, hydrogen can accumulate in the atmosphere. A few hundred bars of hydrogen, which is a small fraction of the likely internal reservoir of hydrogen, are conducive to the formation of a subterranean ocean, which would prevent fracturing of the planetary crust and further outgassing of hydrogen. We further discuss how observations, in conjunction with our model, could tell the history of formation of Planet Nine, and comment on possible CH$_4$/H$_2$ mole ratios in the atmosphere.    

\section{INTRODUCTION}

%{\bf SK revised the introduction some}

There has recently been a great deal of speculation about the possibility of a large planet at a distance of several hundred AU from the Sun \citep{Trujillo2014, Batygin2016, Brown2016, Sheppard2016}.  If this planet forms (i) in the inner solar system and (ii) rapidly enough ($\lesssim$ 1--10~Myr) to capture hydrogen and helium from the protosolar nebula, it will probably have a composition similar to the ice giants Uranus and Neptune. Deriving the cooling history and predictions for the atmospheric structure may then rely on extensions of models developed for ice giants \citep{Linder2016, Ginzburg2016, Fortney2016}.  

In an alternative model, an icy Planet Nine forms {\it in situ} \citep{Kenyon2016}. For the long growth times expected ($\gtrsim$ 0.1--0.3~Gyr), Planet Nine is then a volatile-rich super-Earth composed solely of icy planetesimals. With negligible hydrogen or helium accreted from the protosolar nebula, any atmosphere would have to be outgassed from the interior.  

Predicting the atmospheric structure of an icy Planet Nine requires a new model for the cooling history.
Studies of water-rich super-Earths show that the identity of the chemical species that can outgas into the atmosphere and their corresponding fluxes strongly depend on (i) the high and low pressure crystal structures formed in the mantle, (ii) solubility both in the liquid and solid phases, (iii) the details of the mantle dynamics, and (iv) surface-atmosphere interaction \citep{Levi2013, Levi2014}.  

The purpose of this paper is to construct a basic model for the atmosphere of an icy super-Earth at 250--750~AU from a solar-type star.  Developing a self-consistent model that follows the change in structure of the planet as it accretes and evolves  is a major undertaking beyond the scope of this study.  Instead, we break the problem up into smaller, more manageable pieces, each of which is designed to deal with a different aspect of the problem.  In section $2$, we summarize results for the model of \cite{Kenyon2016} which establishes the accretion rate expected for this case.  In section $3$ we apply this accretion rate to estimate the temperature profile inside the planet using a model adapted from \cite{malamud2016}.  The high temperatures derived in this calculation imply that most of the planet differentiates to a rock core and an ice shell.  With this result established, we use a structure model adapted from the work of \cite{Helled2015}, which uses more careful equations of state and explicitly allows for the pressure dissociation of methane and the release of hydrogen.  This analysis yields an estimate of the hydrogen reservoir in the planet. In section $4$, we quantify the abundance of hydrogen in a secondary outgassed atmosphere surrounding a water-rich super-Earth. 

This analysis suggests plausible outcomes ranging from a thin atmosphere with a pressure of less than 1 bar to a thick atmosphere with a pressure of several hundred bars.  In section $5$ we compare this range of structures to those expected for an ice giant, consider the expected abundance ratio of methane to hydrogen, discuss observations which can distinguish between the two possibilities, and outline possible improvements to our approach. We conclude in section $6$ with a brief summary.

\section{SUPER-EARTH MASS PLANETS AT 250--750 AU}

\citet{Kenyon2016} and \citet{bk2016} outline options for placing a super-Earth mass planet in the outer
solar system \citep[see also][]{stern2005,li2016,mustill2016}. 
In the simplest model, the protosolar nebula produces five gas giant planets at 5--20~AU. Gravitational
interactions among these gas giants scatter one into the outer solar system. If a massive 
protosolar nebula extends to 200--500~AU, dynamical friction between the scattered planet and the gaseous
nebula circularizes the orbit of the planet at 300--600~AU \citep{bk2016}. As a variant of this model, 
physical processes in another planetary system produce a low mass gas giant orbiting at several hundred 
AU. A close encounter between the solar system and this other planetary system allows the Sun to capture 
this planet on an eccentric orbit at 300--600~AU \citep{kb2004d,li2016,mustill2016}. Although it is much
more likely that either of these mechanisms results in an ice giant on a wide orbit, it is possible 
that the planet could be scattered or captured before it accretes hydrogen and helium from the nebula. 
The planet is then a large ball of ice and rock.

\citet{kb2015a,Kenyon2016} explore models where massive planets grow {\it in situ} 
from a massive ring of icy solid material leftover from the formation of the solar system
\citep[see also][]{stern2005,kb2008}. If the ring is composed of a mono-disperse swarm of 
1~cm to 100~cm objects, the time for agglomeration processes to produce a super-Earth mass 
planet is \citep{kb2015a}:
\begin{equation}
t_{SE1} \approx 1 
\left ( \frac{15~\mearth}{M_0 } \right )
\left ( {a \over {\rm 125~AU} } \right )^{3/2} ~ {\rm Gyr} ~ ,
\label{eq: timeSE1}
\end{equation}
where $M_0$ is the initial mass in solids and $a$ is the orbital distance from the Sun.
Within the 4.5~Gyr lifetime of the solar system, this process allows growth of massive
icy super-Earths at 100--300~AU. Growth of icy planets beyond 300~AU requires a very 
massive ring or a formation time scale much longer than 4.5~Gyr.

\begin{figure}[ht]
\centering
\includegraphics[scale=0.80]{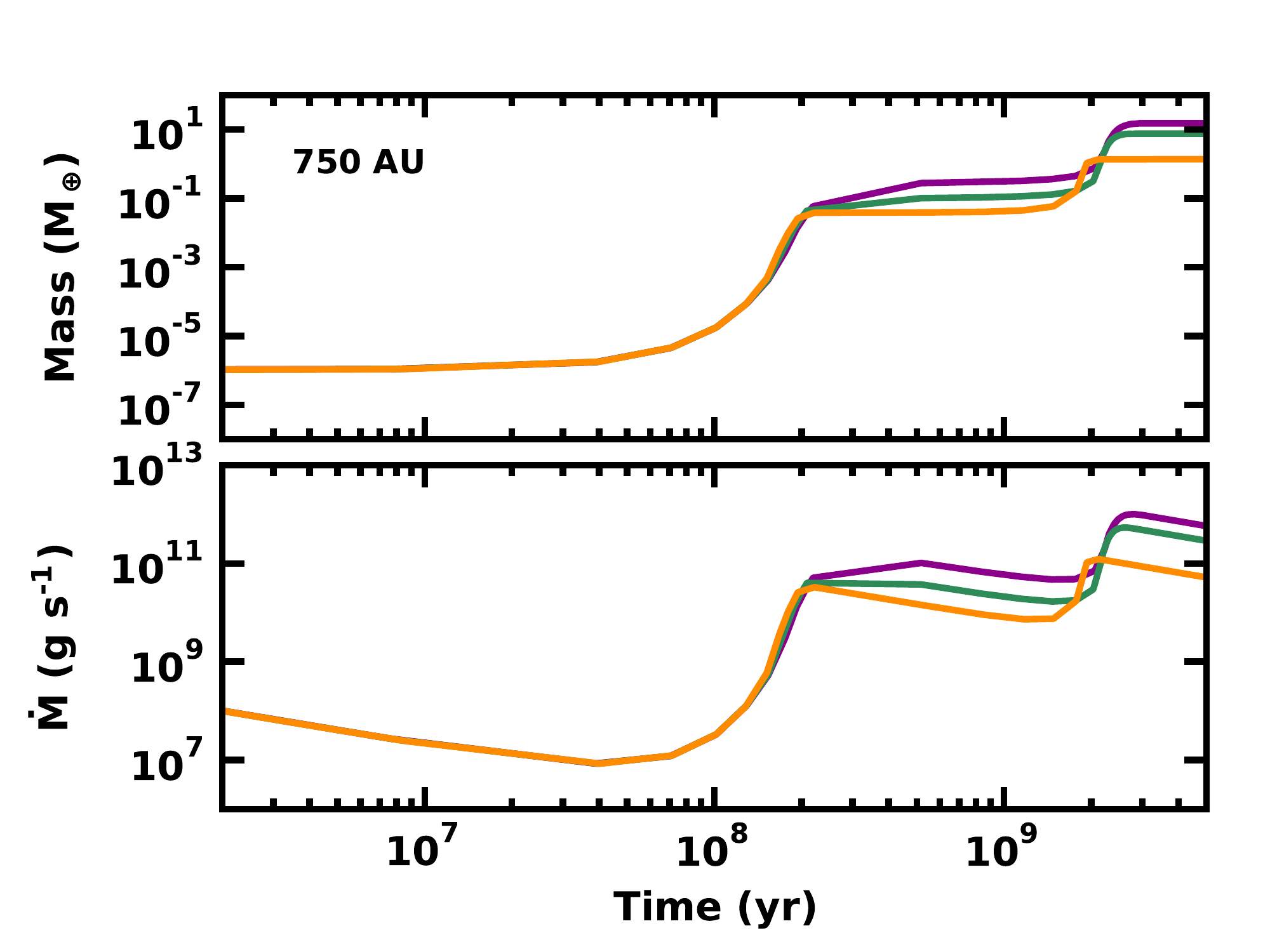}
\caption{\footnotesize{Evolution of icy planets at 750~AU. {\it Upper panel}: Mass of the largest object as a function of time for calculations with one (violet curve), two (green curve), and four (orange curve) 100~km objects accreting from a sea of 1~cm pebbles. {\it Lower panel:} As in the upper panel for the accretion rate onto the largest objects.}}
\label{fig:mmdot}
\end{figure}

If the ring of icy pebbles cools dynamically, a gravitational instability can produce 
one or more protoplanets with radii of 100~km or larger \citep[][and references 
therein]{Kenyon2016}.  Fig.~\ref{fig:mmdot} summarizes the evolution of these protoplanets 
accreting from a ring of icy 1~cm pebbles at 750~AU. After roughly 100~Myr of slow growth, 
the protoplanets undergo a phase of `runaway growth' where their masses increase from 
$10^{-6}$~\mearth\ to roughly 0.1~\mearth.  As they grow in mass, these protoplanets stir 
up the leftover pebbles to higher orbital eccentricity, which slows growth. During the 
next 1--2~Gyr, destructive collisions among the leftovers generates a sea of 0.1--1~mm 
fragments. Over time, collisional damping among the fragments circularizes their orbits.  
Protoplanets can then accrete the fragments rapidly, leading to a second phase of runaway 
growth where protoplanets reach super-Earth masses. 

Numerical calculations with a variety of starting conditions suggest that systems of 1--2 
protoplanets are much more likely to produce 10--15~\mearth\ planets than systems of four 
or more protoplanets. The time scale to reach super-Earth masses is \citep{Kenyon2016}:
\begin{equation}
t_{SE2} \approx 200 
\left ( {15~\mearth \over M_0 } \right )
\left ( {a \over {\rm 250~AU} } \right )^n ~ {\rm Myr} ~ ,
\label{eq: timeSE2}
\end{equation}
where $n \approx$ 2--2.5. This {\it in situ} growth time for protoplanets (100~Myr to 
1--2~Gyr) is much longer than the 5--10~Myr lifetime of the protosolar nebula 
\citep{will2011}.  Thus, these protoplanets never accrete gas from the protosolar
nebula and are simply balls of ice and rock. 

In either the {\it in situ} formation or the scattering scenario, a massive Planet Nine on an orbit with eccentricity $e = 0.1-0.2$ and semimajor axis $a \sim 400-1000$\,AU clears away icy pebbles and planetoids along its orbit over millions of years. Although the architecture of the outer solar system provides some constraints on the mass and orbit of a putative Planet Nine (e.g., \cite{Batygin2016}, \cite{Brown2016}, etc), these results are somewhat controversial \citep{Shankman2016} . If Planet Nine is discovered, the properties of its atmosphere will provide important clues about formation mechanisms (e.g., \cite{Fortney2016}). Thus, we consider constraints based on models where Planet Nine is originally a ball of ice with negligible atmosphere.

\section{SIZE OF THE H$_2$ RESERVOIR}

Planetesimals at 100--1000~AU which agglomerate into an icy Planet Nine are probably composed of a mixture of rock and ice. To derive an estimate of the composition, we adopt the relative elemental abundances of \citet{Lodders2010} and assume that the most abundant rock forming elements, Fe, Mg, and Si, combine with O to form rocky material. Any remaining O then forms water; leftover C and N form CH$_4$ and NH$_3$, respectively.  All of the leftover H forms H$_2$, while the remaining abundant elements, He and Ne remain as atomic noble gases. At the low temperatures ($\sim 20$\,K) relevant to this region of the solar system, CH$_4$, NH$_3$, and H$_2$O are frozen. Thus, any planetesimal is composed of rock plus these three ices.  Experiments show that CH$_4$ under pressure dissociates to more complex hydrocarbons and releases hydrogen.  The exact pressure of dissociation depends on temperature \citep{Kolesnikov2009, Hirai2009, Gao2010, Lobanov2013}.  To estimate the pressures and temperatures inside the planet we need to provide a model of the interior.  

A proper calculation of the interior structure of an icy super-Earth requires (i) an accretion rate, (ii) the energy released by accretion, gravitational compression, and radioactive decay, and (iii) an accurate treatment of material and energy transport throughout the planet.  Developing such a code is beyond the scope of this work. For the purpose of estimating the expected structure and temperatures, we adapted a 1-D evolution code that was originally written for much smaller bodies \citep{prialnik2008, malamud2015, malamud2016}.  This code computes the material and energy transport inside a growing body composed of water ice in a rock matrix that is in hydrostatic equilibrium.  Energy of compression is calculated using a Birch-Murnaghan equation of state with variable coefficients.  Further details can be found in the references cited. This model is not strictly applicable to the situation we are considering since the pressures in the interior are too high to be properly modeled by the Birch-Murnaghan EOS.  In addition, it is likely that the planetesimals accreting onto the planet will have rocky pebbles embedded in an ice matrix rather than the other way around.  Despite these issues, the resulting internal structure and heat release should be similar to that derived from a more detailed calculation, and can be used to posit the initial physical and thermal structure. 

\begin{figure}[ht]
\centering
\includegraphics[scale=0.60]{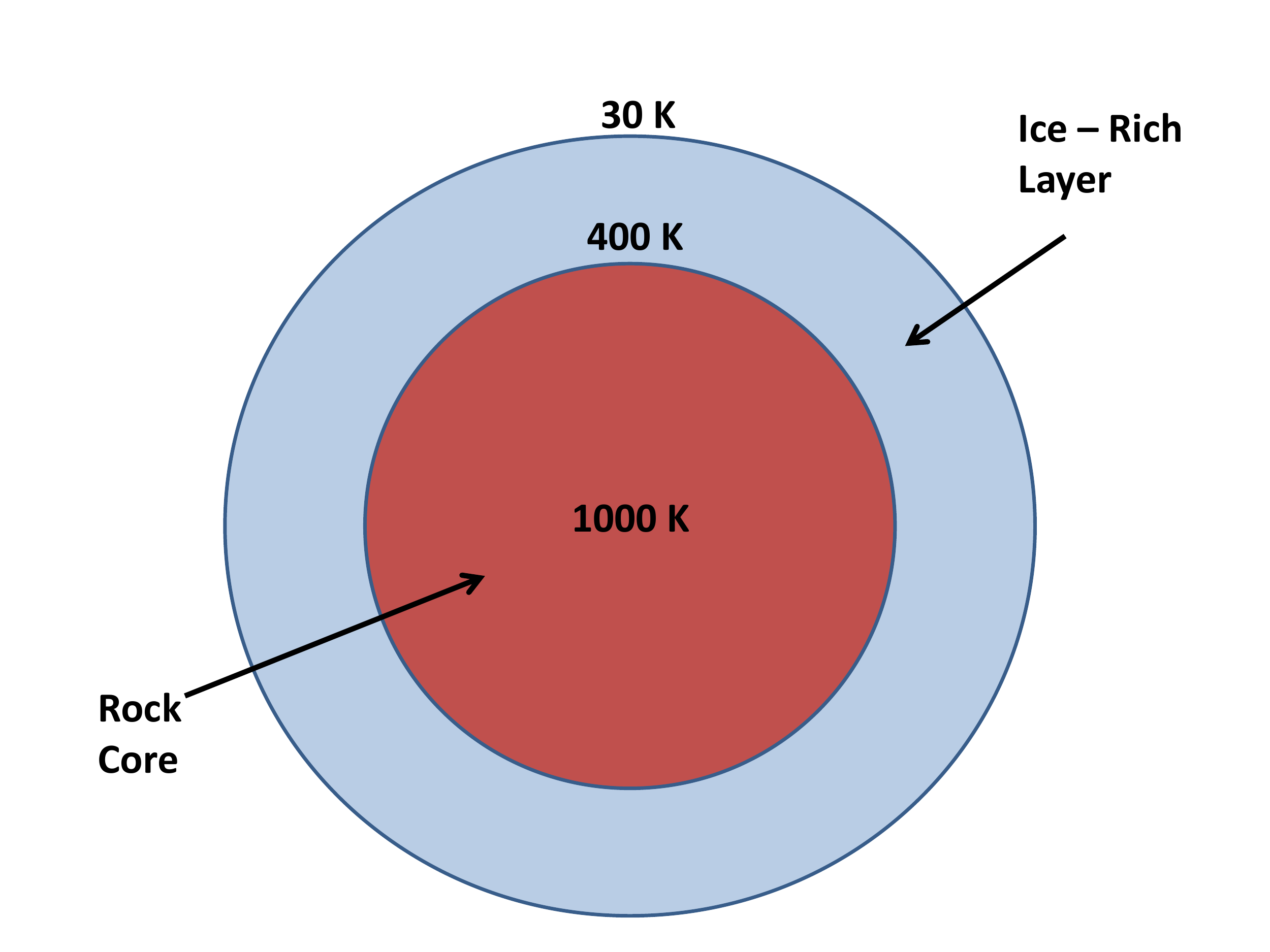}
\caption{\footnotesize{Illustration of the approximate initial physical and thermal structure of the planet.}}
\label{fig:Cartoon}
\end{figure}

For the accretion rate of \cite{Kenyon2016}, the central temperature of the planet rises to $\sim 10^3$\,K.  Under these conditions, most of the interior ice melts and differentiates from the rock, giving a rock core surrounded by a liquid water layer.  Temperatures in the outermost layer are considerably lower, and it does not differentiate completely.  Instead this outermost layer is ice-rich, with a porosity increasing towards the surface.  The details of the temperature structure, thickness of the different layers, porosity, etc. depend on a number of parameter choices, and a qualitative diagram of the assumed initial physical and thermal structure is given in Fig. \ref{fig:Cartoon}. In what follows, we simply assume that the planet is completely differentiated and use the pressures derived from hydrostatic equilibrium to estimate the amount of hydrogen that can be produced from the dissociation of methane trapped in the water ice.  

Another important result from modeling the accretion is that the temperature at the boundary between the rock core and the ice mantle is $\sim 400$\,K.  For such low temperatures, the data of \cite{Gao2010} show that methane dissociates to ethane (C$_2$H$_6$) plus hydrogen above 95\,GPa.  Above 158\,GPa (287\,GPa), methane dissociates to butane C$_4$H$_{10}$ plus hydrogen (carbon/diamond plus hydrogen).  If it can be transported to the surface, the hydrogen formed during this dissociation is volatile enough to form an atmosphere even at the lowest temperatures expected for Planet Nine.

Using the accretion history from section $2$, this analysis suggests that a primordial Planet Nine formed {\it in situ} consists of a rock core surrounded by an ice mantle.  The ice will be a mixture of H$_2$O, CH$_4$, and NH$_3$.  Since we are primarily interested in the behavior of the CH$_4$, we have included the NH$_3$ together with the H$_2$O and used the equation of state of water to represent the mixture.  NH$_3$ is a relatively small fraction of the total, and the abundances of the different species are uncertain, so this approximation seems acceptable.  

To estimate the magnitude of the hydrogen reservoir, we adapted the models of \cite{Helled2015} to compute the pressures inside a planet composed of a rock core surrounded by a mantle with a mixture of H$_2$O and CH$_4$.  Where the pressure is high enough, the CH$_4$ dissociates to ethane, butane or carbon (depending on the pressure) and releases a corresponding amount of H$_2$.  We stress that we do not compute a detailed model of Planet Nine's interior; rather, we estimate the amount of H$_2$ available for outgassing.

\begin{figure}[ht]
\centering
\includegraphics[scale=0.60]{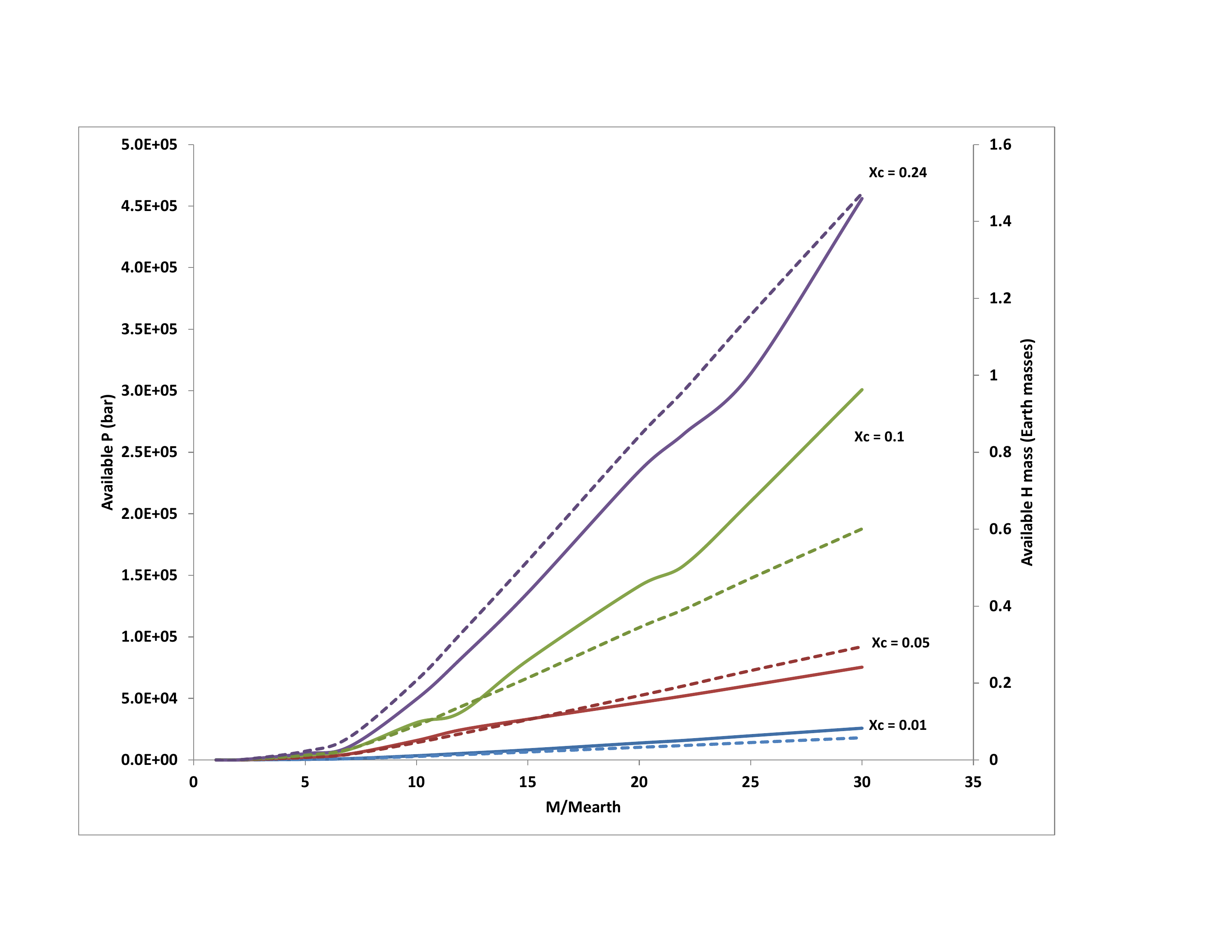}
\caption{\footnotesize{Pressure (solid curves) at the base of the hydrogen atmosphere, and its equivalent mass (dashed curves), as a function of planetary mass for CH$_4$ mass fractions of 0.24 (purple curve), 0.1 (green curve), 0.05 (red curve) and 0.01 (blue curve) if all the available internal hydrogen is assumed to be outgassed.}}
\label{fig:Hpress}
\end{figure}

For the rock core we use the equation of state for SiO$_2$.  The equations of state for SiO$_2$, H$_2$O, CH$_4$, and C$_4$H$_{10}$ are taken from the SESAME tables.  There are no good data for the equation of state of C$_2$H$_6$ at high pressure; experiments by \cite{Goncharov2013} indicate that its density is approximately a factor of 1.2 higher than that of methane over a large range of pressure.  Since C$_2$H$_6$ is a relatively minor component, we modeled it by simply multiplying the density of CH$_4$ by 1.2.  The pressure-density relation inside the planet was calculated assuming constant temperature.  Isotherms of 50\,K and 500\,K gave nearly identical results, so this assumption is reasonable for estimating the hydrogen reservoir.

For solar abundances we expect mass fractions of $X_{H_2O}=.437$, $X_{rock}=.328$, and $X_{CH_4}=.235$.  We have also run cases where the ratio $X_{H_2O}/X_{rock}$ was kept constant but a lower methane abundance was assumed.  We considered $X_{CH_4}=0.1$, 0.05, and 0.01.  The results are shown in Fig.\,\ref{fig:Hpress}.  As can be seen from the figure, a planet with a mass of 5$\mearth$ or more will have high enough pressures in the water-methane layer to produce H$_2$ by CH$_4$ dissociation.  For the case of $X_{CH_4}=0.01$ and a 5$\mearth$ planet, enough hydrogen is produced to provide an atmosphere with a pressure of 430 bar at its base. It should be noted that the \cite{Lodders2010} H$_2$O to rock ratio of 1.33 is significantly lower than the older value of 2.13 derived from the abundance tables of \cite{Anders1989}.  Planets with a higher H$_2$O to rock ratio have higher pressures in the H$_2$O-CH$_4$ mantle, which lead to more methane dissociation and a larger reservoir of H$_2$.  Thus our models provide a lower limit to the size of that reservoir.

\section{CONSTRAINTS ON THE EXTENT OF A SECONDARY OUTGASSED H$_2$ ATMOSPHERE}

If hydrocarbon processing in the deep ice mantle releases vast quantities of H$_2$, building a rich H$_2$ atmosphere requires transport to the outer edge of the planet.  The efficiency of this transport depends on the ability of H$_2$ to become incorporated into the water ice matrix. 

The phase behaviour of the binary H$_2$-H$_2$O system was recently studied, mainly for purposes of examining novel hydrogen storage techniques. Up to a pressure of $0.36$\,GPa a sII clathrate hydrate of H$_2$ is stable, which, when filled to capacity, contains $48$ hydrogen molecules per $136$ water molecules in every unit cell \citep{Lokshin2004}. At $0.36$\,GPa there is a structural transformation to a phase which is stable up to about $0.8$\,GPa. It is referred to as the new phase by \cite{Strobel2011}. The nature of this structure is not yet clear, but X-ray diffraction \citep{Strobel2011} and molecular simulations \citep{Smirnov2013} indicate several possibilities. \cite{Smirnov2013} suggested various structures which were tested by looking at the minimum of the free energy and by examining their stability using classical molecular dynamics. These authors suggested either a unit cell composed of $6$ water molecules and $3$ hydrogen molecules or a unit cell composed of $12$ water molecules and $4$ hydrogen molecules. 

At room temperature and approximately $0.8$\,GPa the H$_2$-H$_2$O system transforms from clathrate to filled ice of phase C$_1$, with $36$ water molecules for every $6$ hydrogen molecules. At $2.3$\,GPa a phase transformation to filled ice C$_2$ occurs with equal abundances of water and hydrogen molecules. This latter phase is stable up to $40$\,GPa at room temperature \citep[see][and references therein]{Strobel2011}. Ab initio simulations for zero temperature suggest that at $38$\,GPa (including a zero point correction) a new phase (C$_3$) with a composition of H$_2$O:2H$_2$ becomes stable and remains stable to $120$\,GPa \citep{Qian2014}. The high stability of C$_3$ is attributed to the very low Bader volume of hydrogen in this structure resulting in a highly dense crystal.    

It is not yet known whether the crystal structures just mentioned are also stable at the temperatures prevailing in the deep ice mantle of a possible Planet Nine at 500--1000~AU. However, the results do suggest that H$_2$ is highly soluble in high pressure water ice structures. Therefore, a non-negligible partition coefficient between any internal H$_2$ reservoir and an overlying high pressure water ice matrix is reasonable. This structure will result in the outward transport of internally produced H$_2$ along with the water ice convection cell.

\begin{figure}[ht]
\centering
\includegraphics[trim=0.15cm 4cm 0.2cm 5cm , scale=0.60, clip]{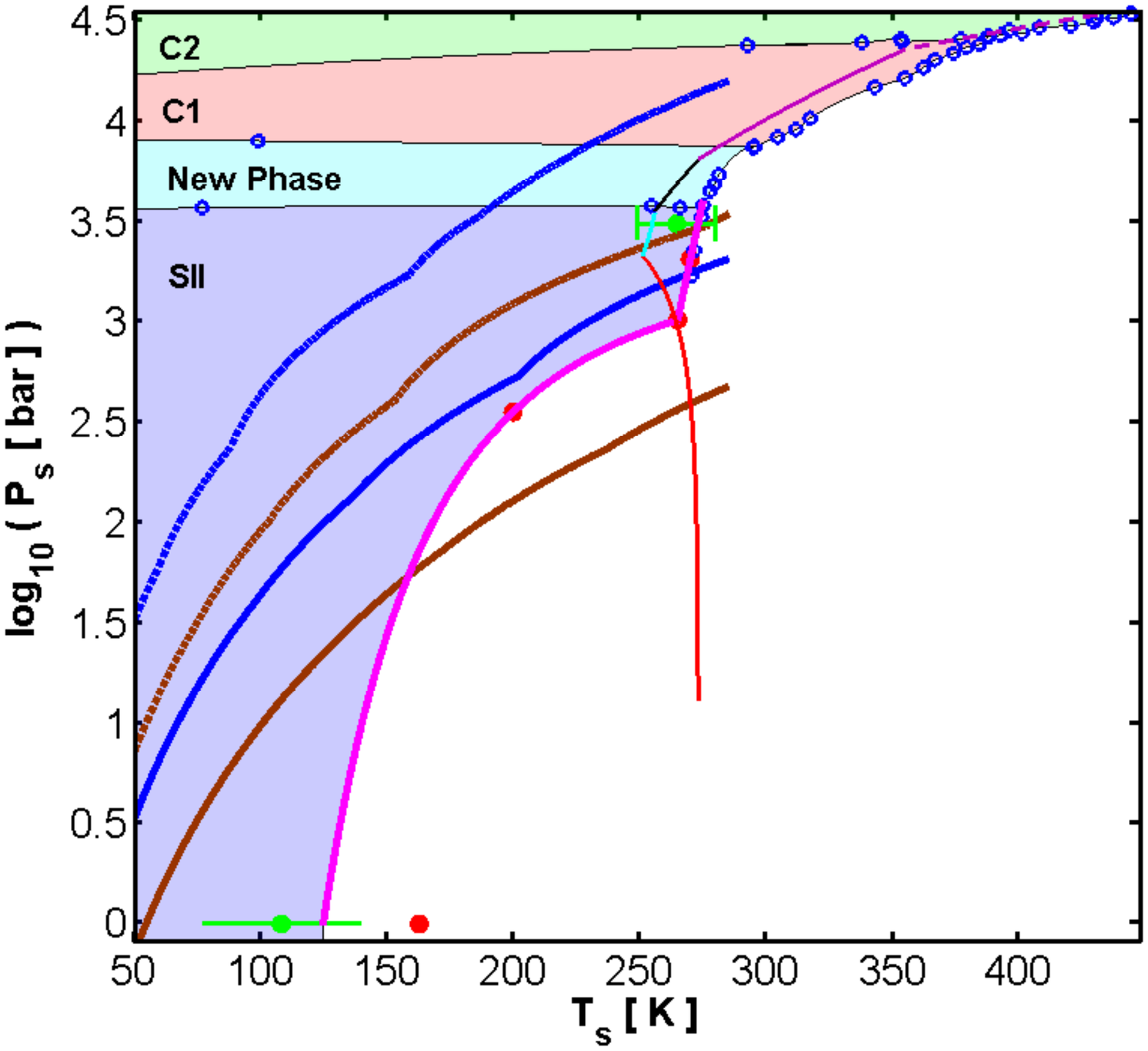}
\caption{\footnotesize{The blue and brown curves are the pressure and temperature relations at the planetary solid surface, i.e. at the base of the H$_2$ atmosphere, for different conditions at unit opacity. Brown curves assume an effective temperature of $30$\,K at unit opacity where the pressure is taken to be either $0.1$\,bar (solid brown curve) or $1$\,bar (dashed brown curve). Blue curves assume an effective temperature of $20$\,K at unit opacity where the pressure is taken to be either $0.1$\,bar (solid blue curve) or $1$\,bar (dashed blue curve). Solid magenta curve is a guide to the eye, going through the published data points for the three phase curves: sII Clathrate-ice Ih-H$_2$ fluid and sII Clathrate-liquid water-H$_2$ fluid. Green data points with their associated errors are from \cite{Mao2004}. Red circles are data points from \cite{Lokshin2006}. Hollow blue circles are data points from \cite{Vos1993}, \cite{Dyadin1999}, \cite{Dyadin1999b}, \cite{Antonov2009} and \cite{Efimchenko2009}. The solid red, cyan, black, purple and dashed purple curves are the pure water ice Ih, ice III, ice V, ice VI and ice VII melt curves, respectively. The different stability fields for the various phases in the H$_2$O-H$_2$ binary solution are represented by the shaded areas.}}
\label{fig:Adiabat}
\end{figure}

We wish to investigate the atmosphere generated from this H$_2$ flux.  To estimate whether such an atmosphere is dynamically stable and how its structure depends on the surface geology and the phase diagram of the binary H$_2$-H$_2$O system, we begin by looking at the phase diagram of sII clathrate hydrate of H$_2$ (see Fig.\,\ref{fig:Adiabat}). The equilibrium of this phase with water ice Ih is difficult to explore experimentally due to the very slow kinetics of formation \citep{Struzhkin2007}. Few experiments trying to identify the three phase curve of this phase with respect to ice Ih and liquid water exist \citep[e.g.,][]{Mao2004,Lokshin2006}.

In Fig.\,\ref{fig:Adiabat} we also plot the relation between the temperature, $T_s$, and pressure, $P_s$, at the base of an H$_2$ atmosphere (see blue and brown curves). To derive this relation, we need to tie the temperature at depth to the temperature at the top of the atmosphere. If we scale the effective temperature of a planet at a distance $a$ from the Sun by 
\begin{equation}
T_{eff}=300\left(\frac{1}{a}\right)^{1/2}
\end{equation}
then the effective temperature of the planet is $T_{eff}\approx 10$\,K at 750~AU and $T_{eff}\approx 20$\,K at 250~AU.  For a hydrogen atmosphere the optical depth is approximately \citep{Lewis1984}
\begin{equation}
\tau=\alpha P^2d\left(\frac{273}{T}\right)^2
\end{equation}
where $P$ is the pressure in bars, $d$ is the thickness of the atmospheric layer in kilometers, which we approximate by the scale height, and $\alpha$ is the thermal IR absorption coefficient for pressure-induced absorption by H$_2$.  With $\alpha\approx 0.1$\,km$^{-1}$amg$^{-2}$, $\tau=1$ corresponds to a pressure of $P=0.1$\,bar, which we take to be the pressure corresponding to the effective temperature\footnote{For comparison, \citet{Fortney2016} adopt $T$(1 bar) = $1.5T^{1.244}_{eff}g^{-0.167}$, from \cite{Guillot1995}, which yields $T$(1 bar) = 1.15-1.45 $T_{eff}$ for a suite of models for 5-50   \mearth objects at $1000$\,AU.}.  Other values suggested in the literature for unit opacity in an H$_2$ atmosphere are: $0.2$\,bar \citep{Wordsworth2012} and around $1$\,bar \citep{Birnbaum1996}. Heat release by radioactive decay could raise the effective temperature to some $30$\,K. Since the optical depth is proportional to $(P/T)^2$, raising the effective temperature by a factor of two would raise the pressure at $\tau=1$ by the same factor. This is well within the range given in the literature.  In Fig.\,\ref{fig:Adiabat} we solve for various unit opacity conditions.   

Once the atmosphere is optically thick, we expect the region below to follow an adiabat \citep[as suggested in][]{Stevenson1999}.  For a pure hydrogen atmosphere this adiabat can be computed from the data in the SESAME tables by integrating the thermodynamic relation $dE=-PdV$ where $E$ is the energy per unit mass of material, and $V$ is the corresponding volume per unit mass. This approach gives us a relation between the surface temperature and pressure. 

From Fig.\,\ref{fig:Adiabat} we are able to identify a \textit{transition point} which depends on the conditions adopted for unit opacity. This point marks the thermodynamic condition where the surface pressure provided by the H$_2$ atmosphere becomes less than the dissociation pressure of the sII clathrate hydrate of H$_2$, if one goes in the direction of increasing temperature. For example, if unit opacity is at $20$\,K and $0.1$\,bar, the transition point is at about $269$\,K and approximately $1720$\,bar. If unit opacity is at $30$\,K and $0.1$\,bar, the transition point is at about $159$\,K and approximately $54$\,bar.

\begin{figure}[ht]
\centering
\includegraphics[trim=0.15cm 4cm 0.2cm 5cm , scale=0.60, clip]{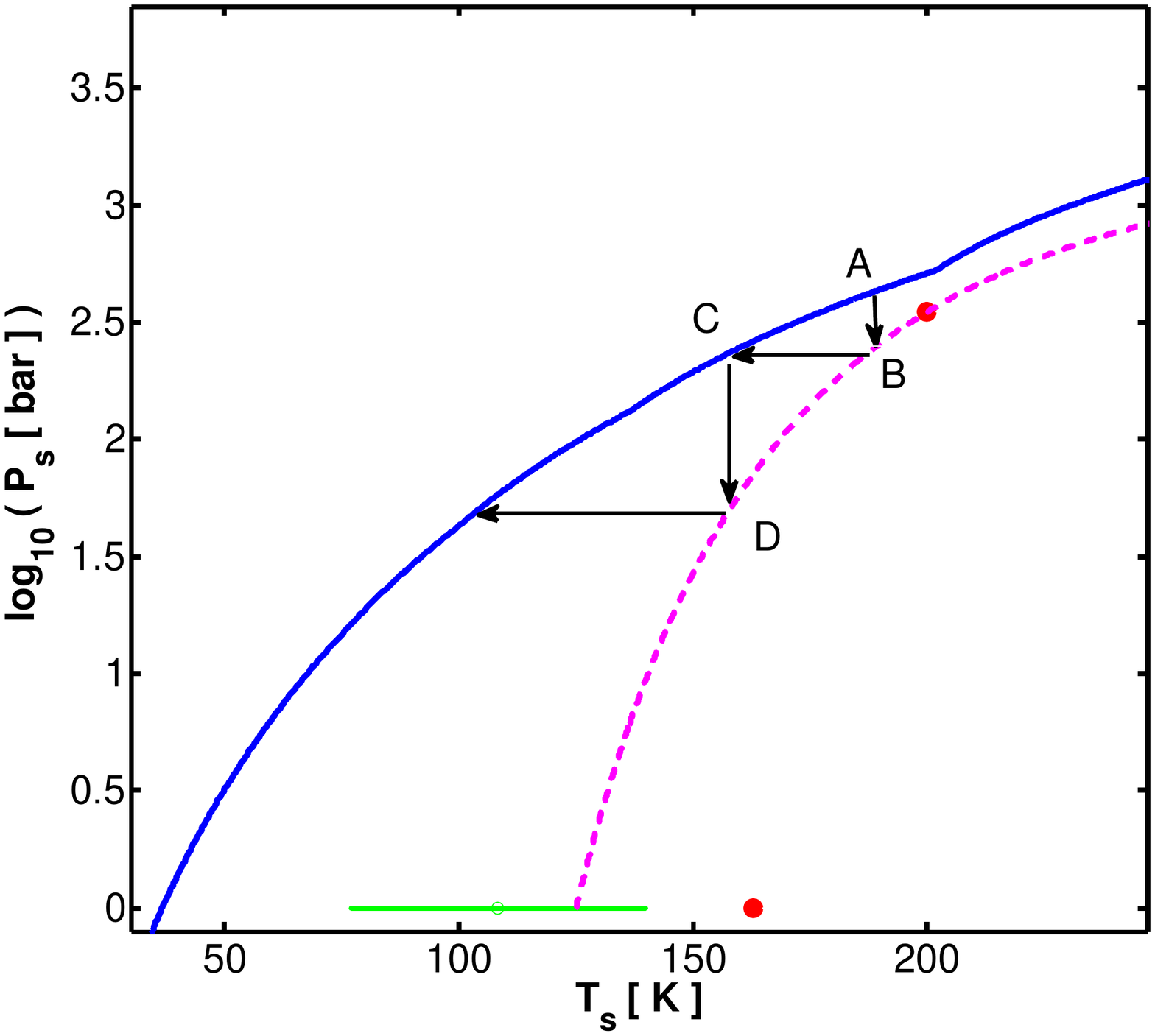}
\caption{\footnotesize{Low temperature section of Fig.\,\ref{fig:Adiabat}. The solid thick blue curve is the pressure and temperature relation at the planetary solid surface, i.e. at the base of the H$_2$ atmosphere, for unit opacity conditions of $20$\,K and $0.1$\,bar. Dashed magenta curve is a guide to the eye, going through the published data points for the three phase curve Hydrate-ice Ih-H$_2$ fluid. Green data point with its associated error is from \cite{Mao2004}. Red circles are data points from \cite{Lokshin2006}. See text for explanation of arrows and letters.}}
\label{fig:RemovalMech}
\end{figure}

This relation has an interesting dynamical consequence. In Fig.\,\ref{fig:RemovalMech} we show a low temperature section of Fig.\,\ref{fig:Adiabat}, to the left of the transition point (for an example adiabat). Let us imagine a water rich planet for which the conditions at the base of its H$_2$ atmosphere fit point A in Fig.\,\ref{fig:RemovalMech}. Since this point is above the dissociation pressure of the sII clathrate hydrate of H$_2$, the hydrogen from the atmosphere reacts with the water ice surface to form clathrate hydrate. The driving force to form this clathrate phase exists until the H$_2$ pressure in the atmosphere reduces to that of point $B$. However, the reduction in the abundance of hydrogen in the atmosphere also cools the surface of the planet, driving the conditions at the base of the atmosphere toward point C. The new H$_2$ pressure at the surface is again higher than the dissociation pressure for the sII clathrate hydrate of H$_2$ for the new and lower surface temperature, and the process continues. 

The lack of low temperature experimental data for the dissociation pressure of sII clathrate hydrate of H$_2$ means we cannot determine at what point this process of atmospheric H$_2$ removal terminates.  However, we may tentatively conclude that there are two dynamically stable end scenarios for the hydrogen atmosphere. One is a poor sub-bar H$_2$ atmosphere if the transition point is not reached, and the other is a H$_2$ atmosphere more massive than the transition point pressure if the system is perturbed or otherwise set at conditions beyond the transition point. We note that this transition point pressure is relatively low compared to what potential internal reservoirs of H$_2$ can provide. In addition, the mechanism for the deposition of the H$_2$ atmosphere into clathrates may be slow. Therefore, even after billions of years such an atmosphere may survive. Finally, it is possible that an outgassing flux of H$_2$ may be established that can counteract this deposition mechanism.

We can estimate the efficiency of the proposed atmospheric H$_2$ removal mechanism. Experiments on the enclathration of CH$_4$ and CO$_2$ show that clathrate formation is initially relatively fast during the stage of a micron-scale surface reaction, and then slows down considerably as clathrate promoters have to diffuse through the outer enclathrated layer in order to reach fresh ice Ih \citep{Staykova2003,Genov2004}. Therefore, the depth from the surface that can participate in the enclathration of atmospheric H$_2$ is:
\begin{equation}
\delta\sim\sqrt{Dt}
\end{equation}  
where $D$ is the diffusion coefficient of H$_2$ in the water matrix, and $t$ is the time that has elapsed. Thus the number of moles of water taking part in this process is:
\begin{equation}
n_{h_2o}\sim\frac{ 4\pi\delta R^2_p\rho_{ice}}{\mu_{H_2O}}
\end{equation}
Here, $R_p$ is the planet's radius, $\rho_{ice}$ is the mass density of water ice Ih and $\mu_{H_2O}$ is the molar weight of water. In a fully occupied sII clathrate hydrate of H$_2$, a unit cell composed of $136$ H$_2$O molecules entraps $48$ H$_2$ molecules \citep{Lokshin2006}. The number of moles of H$_2$ that can be taken out of the atmosphere after time $t$ is:
\begin{equation}
n_{h_2}\sim\frac{48}{136}\frac{ 4\pi\delta R^2_p\rho_{ice}}{\mu_{H_2O}}
\end{equation}  
The relation between the surface pressure due to the H$_2$ atmosphere and the mass, $M_{H_2}$, of this atmosphere is:
\begin{equation}
M_{H_2}g=4\pi R^2_pP_s 
\end{equation} 
where $g$ is the surface acceleration of gravity. After rearranging we have:
\begin{equation}
P_s\sim\frac{48}{136}\frac{\mu_{H_2}}{\mu_{H_2O}}g\rho_{ice}\sqrt{Dt}
\end{equation}
where $\mu_{H_2}$ is the molar weight of diatomic hydrogen.
 
The self-diffusion coefficient of H$_2$ in its hydrate water matrix is the subject of ongoing research \citep[e.g.][]{Nagai2008,Trinh2015}. For our purposes we require the diffusion coefficient at low temperatures where tunneling of H$_2$ is important. \cite{Alavi2007} found that the H$_2$ diffusion coefficient for crossing the hexagonal face of a sII clathrate hydrate large cage is $10^{-4}$\,cm$^2$\,s$^{-1}$, while crossing a pentagonal cage face corresponds to a diffusion coefficient of $10^{-7}$\,cm$^2$\,s$^{-1}$.  We can estimate the effective self-diffusion coefficient by treating the sII clathrate as a multicomponent system, where we assume each cage type is a ``different" component. According to \cite{weertman1975} this results in an average diffusion coefficient of:
\begin{equation}
D = \frac{10^{-7}\times 10^{-4}}{\frac{8}{24}\times 10^{-7}+\frac{16}{24}\times 10^{-4}} \approx 10^{-6.8}\, [cm^2\,s^{-1}]
\end{equation}
where we use the fact that each sII unit cell is composed of $8$ large cages (first component) and $16$ small cages (second component). 

Let us consider a planet with a mass of 15\,\mearth, for which we calculate a gravitational surface acceleration of $g=1570$\,cm\,s$^{-2}$. After $t=1$\,Gyr the removal mechanism can deposit approximately $4$\,bar of atmospheric H$_2$ into clathrates. This can be translated into an average \textit{critical flux}:
\begin{equation}
F_{cr}=\frac{N_{H_2}}{4\pi R^2_p t}\approx 2.5\times 10^{10} \,[molecules \,cm^{-2}\,s^{-1}]
\end{equation} 
Here $N_{H_2}$ is the number of H$_2$ molecules equivalent to the $4$\,bar deposited in $t=1$\,Gyr.
For the planets we model, if they are below the transition point, having an H$_2$ outgassing flux higher (lower) than the critical flux means their atmospheres would become richer (poorer) in H$_2$ with time.

As mentioned in the introduction, we consider the case of a water-rich super-Earth forming at large heliocentric distances.    
The slow accretion rate \citep{Kenyon2016}, means that it will not reach a mass large enough to capture an H/He envelope from the surrounding nebula before that nebula dissipates. If our water-rich super-Earth experiences a subcritical H$_2$ outflux, it probably has a negligible H$_2$ atmosphere. Clearly, this scenario results in a distinctive mass density and atmospheric composition, different than those of a classical ice giant similar to Neptune or Uranus.  However, a supercritical outgassing flux of H$_2$ enriches the atmosphere with time. To learn whether this evolution leads to a structure similar to an ice giant, we now examine the thermal profile and dynamics of the planetary ice crust and upper mantle, and how these evolve as the atmosphere is enriched with H$_2$. 

In \cite{Levi2014} we have developed a model for the planetary crust and underlying convection for a water-rich super-Earth. In this model we solve for the case of a stagnant lid and a small viscosity contrast. We truncate the thermal conductive profile assumed for the crust by looking for the depth where the Rayleigh number reaches its critical value. Then we follow an adiabat into the planet. We refer the reader to \cite{Levi2014} for a thorough explanation of the technique used. Here we will only discuss the changes we have made to this model in its current application.            

To be consistent with the analysis of the previous section we consider the crust and upper mantle to have an ice composition of H$_2$O and CH$_4$,
with a mole ratio:
\begin{equation}
\eta = \frac{\tilde{n}_{CH_4}}{\tilde{n}_{H_2O}}
\end{equation}
The CH$_4$ forms a sI clathrate hydrate, for which the H$_2$O to CH$_4$ mole ratio is $\xi=5.75$, assuming full occupancy of the clathrate cages. The excess H$_2$O forms pure Ih water ice. This view is corroborated by experiments following the succession of phases with pressure in the H$_2$O-CH$_4$ system \citep{Hirai2001,Loveday2001,Ohtani2010}. The fraction of the ice volume occupied by the sI CH$_4$ clathrate hydrate is:
\begin{equation}
\phi=\frac{V_{ns}}{V_{ns}+V_{H_2O}}
\end{equation} 
where $V_{ns}$ is the total volume of the sI CH$_4$ clathrate hydrate (i.e the non-stoichiometric crystal) in the upper ice mantle and $V_{H_2O}$ is the total volume of the pure water ice Ih. After several algebraic steps one can show that:
\begin{equation}
\frac{1}{\phi}=1+\left[1+\frac{1}{\xi}\frac{\mu_{CH_4}}{\mu_{H_2O}}\right]^{-1}\frac{\rho_{ns}}{\rho_{H_2O}}\left(\frac{1-\eta\xi}{\eta\xi}\right)
\end{equation}  
Here $\mu_{CH_4}$ and $\mu_{H2O}$ are the molar masses of CH$_4$ and H$_2$O respectively. The mass densities of the appropriate pure water ice and non-stoichiometric phases are $\rho_{H_2O}$ and $\rho_{ns}$ respectively.

The composite mass density is:
\begin{equation}
\rho_{comp}=\rho_{H_2O}\frac{1-\phi}{1-\eta\xi}\left[\eta\frac{\mu_{CH_4}}{\mu_{H_2O}}+1\right]
\end{equation}
We approximate the isobaric heat capacity of the composite using the rule of mixtures:
\begin{equation}
C^{comp}_p=C^{ns}_p\phi + C^{H_2O}_p(1-\phi)
\end{equation}
where $C^{ns}_p$ and $C^{H_2O}_p$ are the isobaric heat capacities of the non-stoichiometric sI CH$_4$ clathrate hydrate phase and of water ice Ih, respectively.   

The thermal conductivity of our assumed composite ice mantle is $\kappa_{comp}$. This composite may be approximated as a continuous pure water ice Ih within which small non-stoichiometric sI CH$_4$ clathrate hydrate grains are embedded. The thermal conductivity of such a composite was solved for by Maxwell \citep{Bird2007transport}, and for our case has the following form:
\begin{equation}
\frac{\kappa_{comp}}{\kappa_{H_2O}}=1+\frac{3\phi}{\left(\frac{\kappa_{ns}+2\kappa_{H_2O}}{\kappa_{ns}-\kappa_{H_2O}}\right)-\phi}
\end{equation}
In the last relation $\kappa_{H_2O}$ is the thermal conductivity of water ice Ih. The thermal conductivity of the non-stoichiometric sI CH$_4$ clathrate hydrate phase is $\kappa_{ns}$. Therefore, the thermal diffusivity of the upper ice mantle is:
\begin{equation}
\alpha_{comp}=\frac{\kappa_{comp}}{\rho_{comp}C^{comp}_P}
\end{equation}

The volume thermal expansivity of our composite material is $\chi_{comp}$. There are various suggestions in the literature for relating the composite expansivity to that of the individual phases composing it. The rule of mixtures and Turner's formula confine the composite expansivity from above and below respectively \citep{Schapery1968,Karch2014}:
\begin{equation}
\frac{K_{ns}\chi_{ns}\phi + K_{H_2O}\chi_{H_2O}(1-\phi)}{K_{ns}\phi + K_{H_2O}(1-\phi)}\quad <\quad \chi_{comp}\quad < \quad\chi_{ns}\phi + \chi_{H_2O}(1-\phi)
\end{equation}
where $K_{H_2O}$ is the bulk modulus of water ice Ih, and $K_{ns}$ is the bulk modulus of the non-stoichiometric sI CH$_4$ clathrate hydrate.
We will adopt Turner's formula, making our adiabatic profile steeper. This means more H$_2$ is required in the atmosphere in order to increase the surface temperature, if the adiabat is to cross the melt curve of ice Ih.

The thermodynamic data for the sI CH$_4$ clathrate hydrate phase was given in detail in \cite{Levi2014}. The mass density, isobaric heat capacity and volume thermal expansivity for water ice Ih are taken from \cite{Feistel2006}. The thermal conductivity of water ice Ih is taken from \cite{slack1980}. The bulk modulus for water ice Ih is taken from \cite{Helgerud2009}.

\begin{figure}[ht]
\centering
\includegraphics[trim=0.15cm 4cm 0.2cm 5cm , scale=0.60, clip]{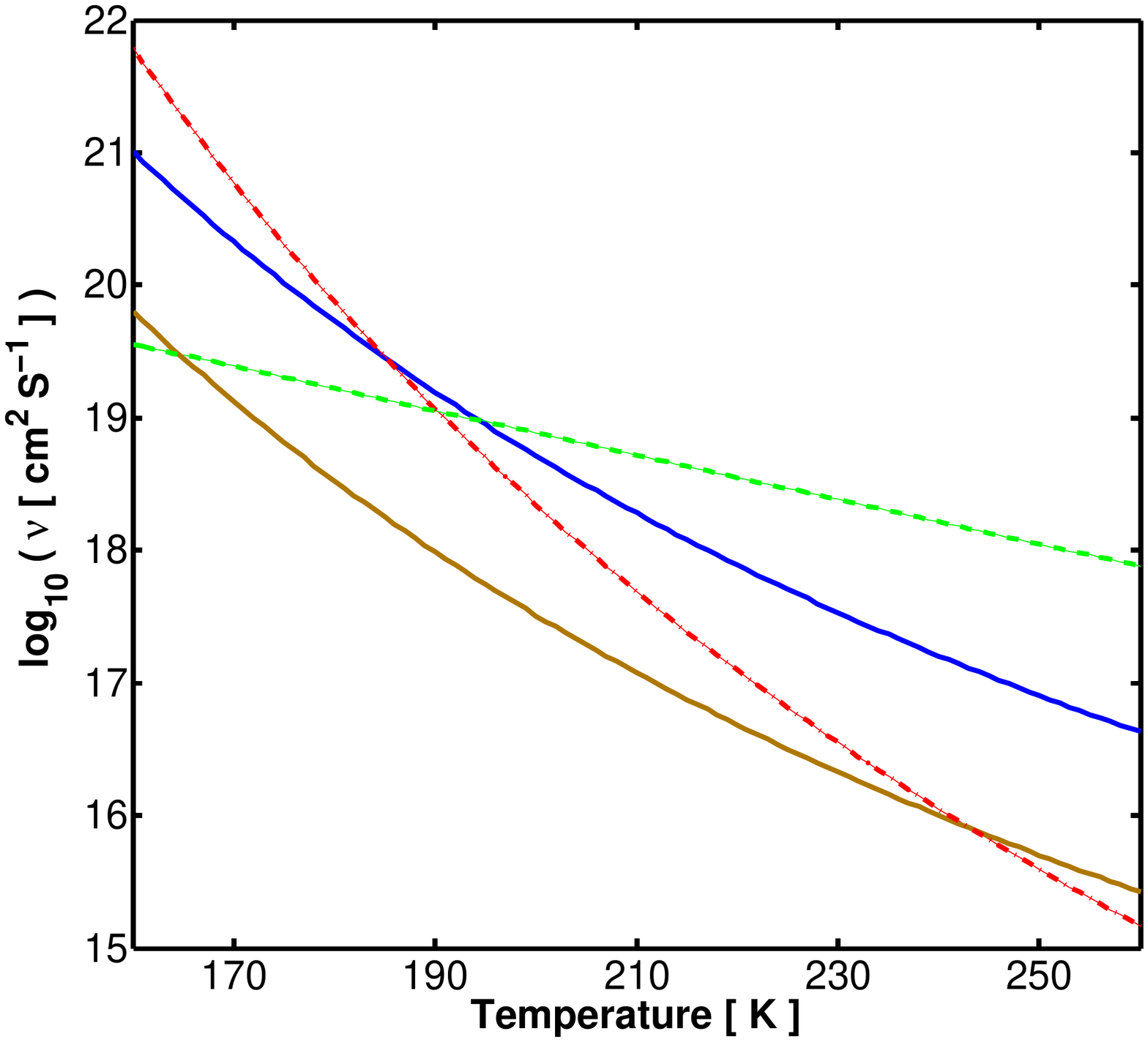}
\caption{\footnotesize{Kinematic viscosity at a reference pressure of $1000$\,bar. Dashed green curve is the dislocation creep model suggested by \cite{Durham2001} for water ice Ih, for planetary type strain rates. Dashed-dotted red curve is the average model for the viscosity of water ice Ih from \cite{spohn2003}. Solid blue curve is the Newtonian viscosity for sI CH$_4$ clathrate hydrate from \cite{Levi2014} for grain diameters of $800\,\mu$m. Solid brown is the same model but assuming grain diameters of $200\,\mu$m.}}
\label{fig:viscosity}
\end{figure}

Lastly, we discuss the choice we have made for the viscosity. \cite{Durham2001} suggested values for the effective viscosity due to dislocation creep, for planetary type strain rates. These are an extrapolation over several orders of magnitude in the strain rate. In \cite{Levi2014} we have suggested a model for the Newtonian creep in sI CH$_4$ clathrate hydrate, which is a function of the grain size. In Fig.\,\ref{fig:viscosity} we compare between different models for the viscosity. Above approximately $200$\,K, it seems that dislocation creep is not as active as diffusion creep. Also, varying the grain diameters in our diffusion creep model \citep[see][]{Levi2014}, between $200-800\,\mu$m spans reasonable values for the viscosity in our ice crust and upper mantle, when compared with the other viscosity models.  

\begin{figure}[ht]
\centering
\includegraphics[trim=0.15cm 19cm 0.2cm 3cm , scale=0.80, clip]{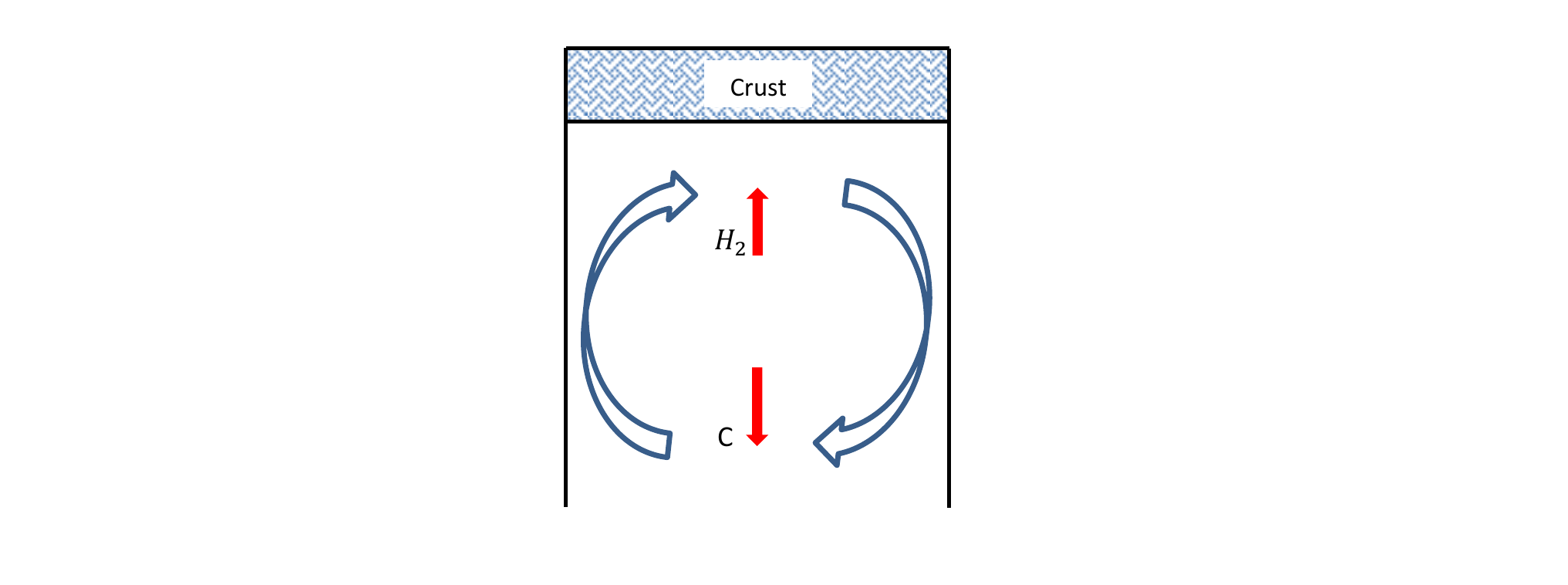}
\caption{\footnotesize{The planetary crust overlies a large-scale convection cell, capable of breaking the crust. In the convection cell CH$_4$ is dragged inward, where under high pressures it dissociates and releases H$_2$. This hydrogen may be transported outward and reach the atmosphere. See also type III planet in \cite{Fu10}.}}
\label{fig:typeIII}
\end{figure}

Again, we assume a planet with a mass of 15\,\mearth and a surface gravitational acceleration of $1570$\,cm\,s$^{-2}$. With the rock mass fraction given in the previous section we can estimate the heat released due to radioactive decay \citep[see Eq.\,(23) in][]{Levi2014}.  This gives a surface heat flux of $46$\,erg\,cm$^{-2}$\,s$^{-1}$.  Starting from an estimated surface temperature of $20-30$\,K, we find our planet will be of type III of the four planetary types defined by \cite{Fu10} (see illustration in Fig.\,\ref{fig:typeIII}). 

\begin{figure}[ht]
\centering
\includegraphics[trim=0.15cm 19cm 0.2cm 3cm , scale=0.80, clip]{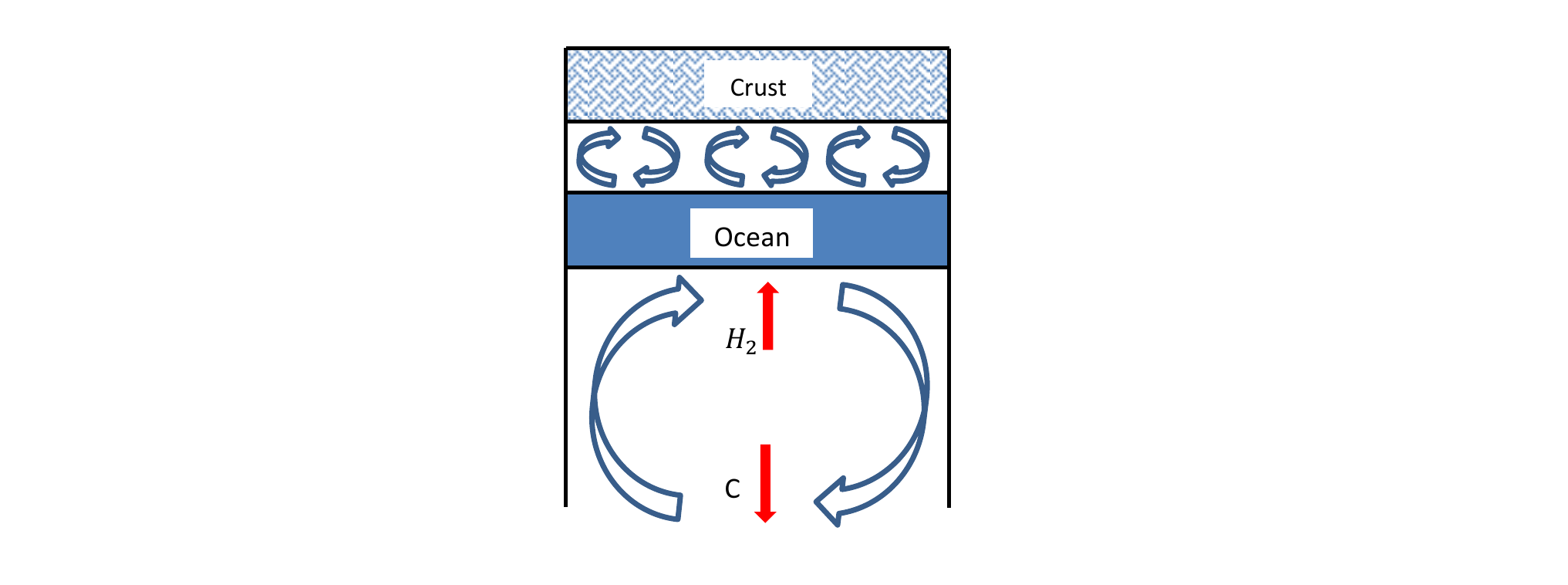}
\caption{\footnotesize{The planetary crust terminates with the initiation of small-scale convection, overlying a subterranean ocean. The small-scale convection cannot break the crust apart. Therefore, hindering any outgassing of H$_2$ out into the atmosphere. See also type II planet in \cite{Fu10}. }}
\label{fig:typeII}
\end{figure}

Since the surface temperature is lower than the melt temperature for water ice Ih, a solid crust forms. As discussed in \cite{Fu10}, for an ice mantle with no asthenosphere, and in \cite{Levi2014} for an ice mantle with an asthenosphere, a non-partitioned ice mantle convection cell would likely apply a high enough stress at the base of the crust to break it apart.
Breaking the planetary crust exposes fresh deep ice to the surface conditions. This should result in the outgassing of H$_2$. If H$_2$ outgassing flux is kept \textit{supercritical}, the atmosphere would become richer in hydrogen with time, thus increasing the temperature at the outer solid surface of the planet, $T_s$. This increase in the surface temperature shifts the temperatures along the planetary crust and upper mantle to higher values as well. Increasing the surface temperature enough may drive the deep thermal profile across the melt curve of water ice Ih, forming a subterranean ocean. The planet now attains a type II structure (see illustration in Fig.\,\ref{fig:typeII}). This however depends on the conditions at unit opacity in the H$_2$ atmosphere.

\begin{figure}[ht]
\centering
\includegraphics[trim=0.15cm 4cm 0.15cm 4cm , scale=0.60, clip]{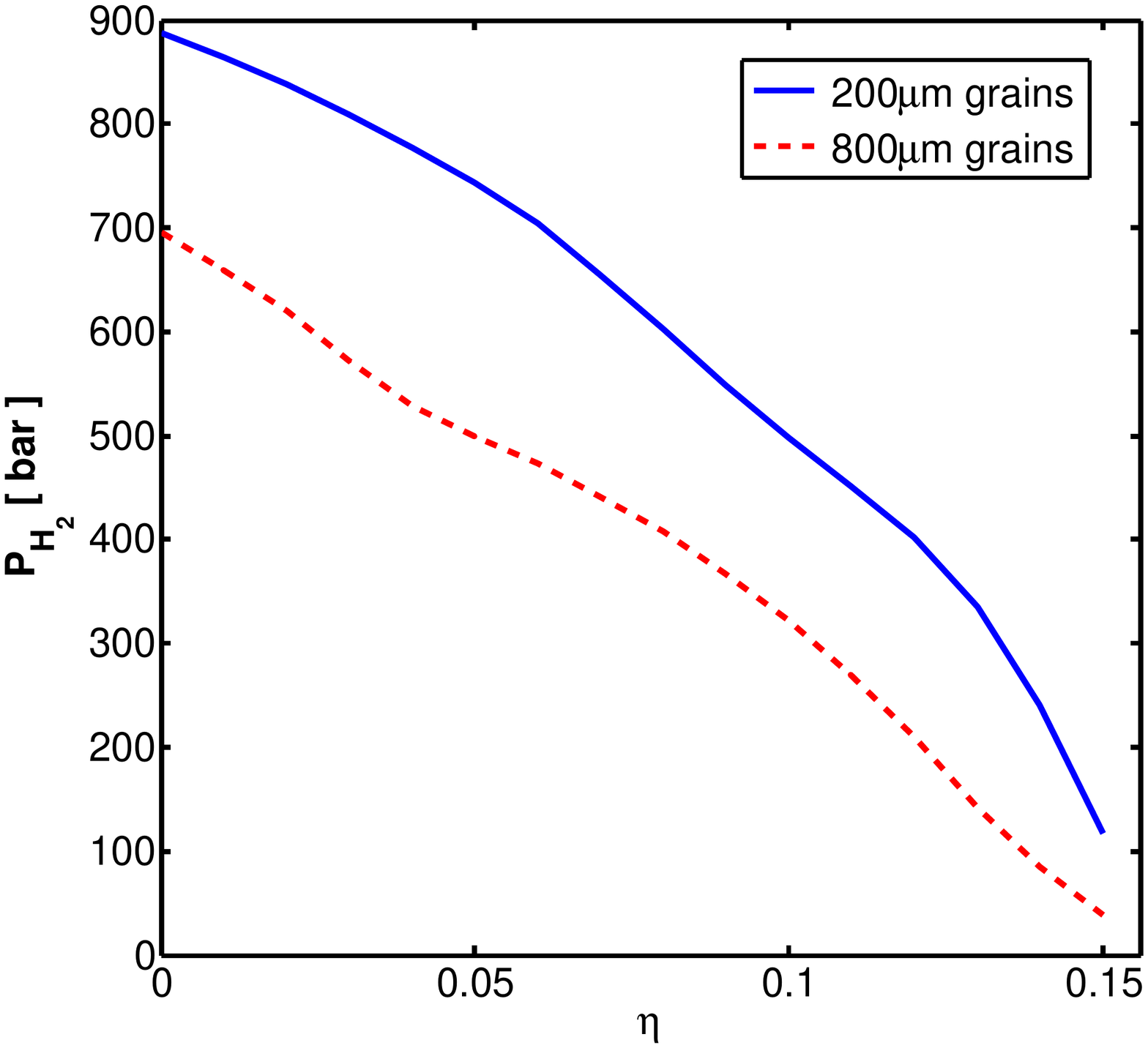}
\caption{\footnotesize{The minimal pressure of atmospheric H$_2$ required to transform a type III stratified water-rich superEarth into a type II stratified planet. $\eta$ is the CH$_4$ to H$_2$O mole ratio of our assumed ice matrix. The solid blue curve is for our viscosity model assuming a grain size of $200\mu$m, and the dashed red curve is for a grain size of $800\mu$m. Unit opacity conditions in the atmosphere are taken here to be $0.1$\,bar with an effective temperature of $20$\,K.}}
\label{fig:H2Pressure}
\end{figure}

In Fig.\,\ref{fig:H2Pressure} we plot the minimum amount of H$_2$ in the atmosphere required to shift the planet into a type II stratification model. As an example for the system's behaviour we assume an effective temperature of $20$\,K and a pressure of $0.1$\,bar for the unit opacity conditions in the atmosphere. Assuming an ice matrix richer in CH$_4$ (i.e. larger $\eta$) means the composite has more sI CH$_4$ clathrate hydrate. Because the thermal conductivity of clathrates is lower than that for ice Ih, the more clathrates in the composite, the higher the temperature increase along the crust, before reaching convective instability. Therefore, a lower surface temperature (i.e. less H$_2$ in the atmosphere) is sufficient to transform the planet into a type II water-rich planet. In addition, the larger grain size results in a higher viscosity and thus a thicker crust. In the crust the temperatures rise steeply with depth, hence a thicker crust means that less H$_2$ in the atmosphere is needed in order to cross the melt curve of ice Ih.

For the type II planetary stratification of Fig.\,\ref{fig:typeII} we estimate the stress applied by the small scale convective layer on the overlying crust in the following way: The maximum velocity in the small scale convection cell is approximately:
\begin{equation}
u_{max}\approx 0.271\frac{\alpha_{comp}}{d}Ra^{2/3}
\end{equation}
where $Ra$ is the Rayleigh number and $d$ the thickness of the small scale convective layer, between the crust and the subterranean ocean.
The basal stress is:
\begin{equation}
\tau_b\approx \nu\rho_{comp}\frac{u_{max}}{d}
\end{equation}
where $\nu$ is the kinematic viscosity. The stress acting to fracture the crust is thus:
\begin{equation}
\sigma\approx \tau_b\frac{d}{d_{cr}}
\end{equation}  
where $d_{cr}$ is the thickness of the crust. 

We find that increasing $\eta$ from $0$ to $0.15$ results in $d$ increasing from $7$\,km to $11$\,km. The crust in general has a thickness ranging from $1.5$\,km to $3$\,km. The Rayleigh number changes by two orders of magnitude over this range of $\eta$. The stress $\sigma$ falls in the range of $10^{-2}$ - $10^{-1}$\,MPa. This is less than the tensile strength of ice Ih \citep[$1.8$\,MPa at $233$\,K,][]{hobbs} and for sI CH$_4$ clathrate hydrate \citep[$0.2$\,MPa at room temperature,][]{Jung2011}. Therefore, the crust probably cannot be broken apart by the underlying small scale convection. As a consequence any further outgassing of internal H$_2$ will be hindered, and the values for the pressure of atmospheric H$_2$ given in Fig.\,\ref{fig:H2Pressure} represent the likely maximum values.

\begin{figure}[ht]
\centering
\mbox{\subfigure{\includegraphics[width=7cm]{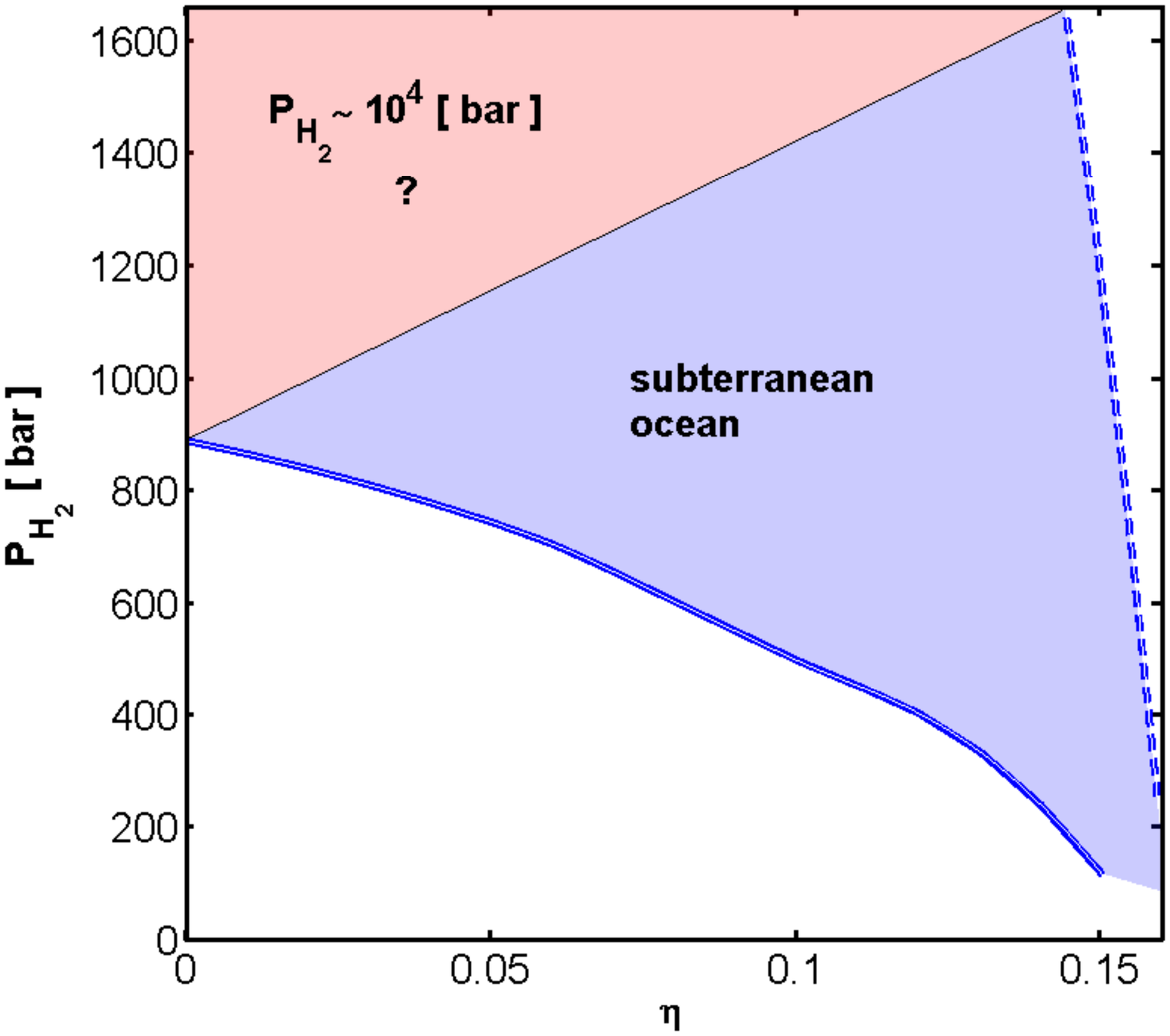}}\quad \subfigure{\includegraphics[width=7cm]{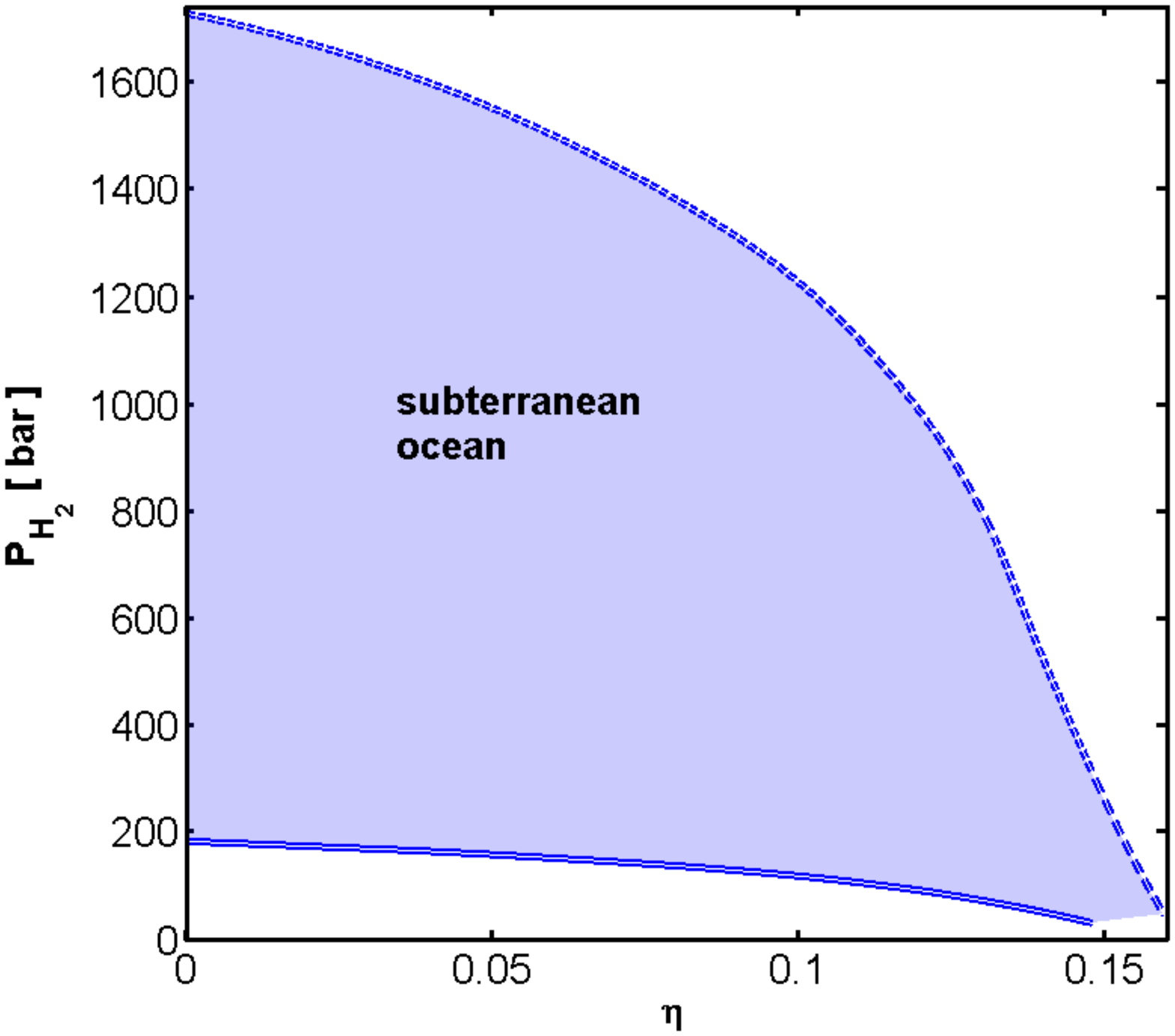}}}
\caption{\footnotesize{Here we plot the effect of varying the pressure at unit opacity for a pure H$_2$ atmosphere, assuming an effective temperature of $20$\,K (\textbf{left panel}) and an effective temperature of $30$\,K (\textbf{right panel}). Thick solid blue curves are for a unit opacity pressure of $0.1$\,bar. For this value of the unit opacity pressure, for every crustal composition, from pure water ($\eta=0$) to pure SI CH$_4$ clathrate hydrate ($\eta=0.174$) a subterranean ocean forms, truncating any further outgassing of H$_2$ into the atmosphere. For the effective temperature of $20$\,K this curve is also the thick solid blue curve in Fig.\ref{fig:H2Pressure}. For higher values of the unit opacity pressure (e.g. $1$\,bar, thick dashed blue curves) and for an effective temperature of $20$\,K, there will be a value for $\eta$ below which it will no longer be possible to form a subterranean ocean. As a result, for a supercritical flux of H$_2$, the internal reservoir of H$_2$ may outgas into the atmosphere.}}
\label{fig:ConstrainTransition}
\end{figure}

The ability to transition into a type II planet depends on the composition of the crust (i.e. $\eta$) and the conditions adopted for unit opacity in the atmosphere. The likely formation of a subterranean ocean may be attributed to the fact that, while the temperature increases inward along the conductive crust the melt temperature of ice Ih decreases with increasing pressure. For lower effective temperatures or higher unit opacity pressures, conditions at the bottom of an H$_2$ atmosphere can miss the triple point of ice Ih-ice III-liquid water. Beyond this triple point the melt temperature of high pressure ice polymorphs increases with increasing pressure, thus rendering the formation of a subterranean ocean less likely. In such a case the planet remains a type III planet, maintaining the ability to outgas internal H$_2$, and possibly forming a rich H$_2$ atmosphere on the order of $10^4$\,bar. This behaviour is shown in the left panel in Fig.\ref{fig:ConstrainTransition}, where we used $T_{eff}=20$\,K as an example, and varied the unit opacity pressure in the range of $0.1$\,bar to $1$\,bar. As is seen in the figure, having less CH$_4$ in the crust lessens the probability for a subterranean ocean, and the ability of the atmosphere to isolate itself from the planetary inner workings.

In the right panel in Fig.\ref{fig:ConstrainTransition}, we have adopted $T_{eff}=30$\,K as an example, and varied the unit opacity pressure in the range of $0.1$\,bar to $1$\,bar. As is seen in this panel the conditions at the bottom of the H$_2$ atmosphere never miss the triple point of ice Ih-ice III-liquid water, and thus a subterranean ocean always forms. For this higher effective temperature the internal reservoir of H$_2$ cannot outgas completely into the atmosphere. The unbreakable crust truncates the atmosphere at a value of no more than $\sim 1000$\,bar base pressure of H$_2$.

\section{DISCUSSION}

Our analysis suggests that the pressure in the mantle of an icy super-Earth formed {\it in situ} at 250--750~AU is sufficient to liberate H$_2$ from methane. Radial transport of this H$_2$ yields an outgassed atmosphere with a pressure of $\lesssim$ 1~bar to $\gtrsim$ 100~bar. Even if the entire hydrogen reservoir of $\sim 10^4$\,bar is outgassed, however, the mean density of a 15~$\mearth$ icy super-Earth has a fairly small range, 2.4--3.0\,g\,cm$^{-3}$.  Reducing the surface pressure of hydrogen by two orders of magnitude decreases the thickness of the atmosphere by five scale heights ($\sim 100-500$\,km depending on the atmospheric temperatures) without substantially changing the radius of the bulk of the planet and increasing the expected mean density by less than ten percent. Despite the potentially large atmosphere, the mean density is significantly larger than the mean density of Uranus (1.27\,g\,cm$^{-3}$) or Neptune (1.64\,g\,cm$^{-3}$). Thus, it is possible to distinguish a Planet Nine formed {\it in situ} from one which accreted gas from the protosolar nebula.

Deriving the true underlying structure of a real Planet Nine beyond 250~AU requires measurements of the mass and radius.  Two approaches can place limits on the radius: (i) fits of model atmospheres \citep[e.g.,][]{Fortney2016} to multi-wavelength observations of the emitted flux and (ii) occultations of background stars. Estimates for the mass rely on (i) derivation of log~$g$ from the model atmosphere or (ii) identification of a satellite and direct measurement of its period and semimajor axis. 

Although straightforward, these measurements may be challenging. Measuring the spectral energy distribution beyond 2~$\mu m$ requires a Planet Nine that is bright enough for detection with {\it JWST}. Occultation observations need a Planet Nine within a field sufficiently dense with reasonably bright stars. Based on HST observations of Pluto \citep[e.g.,][and references therein]{Brozovic2015}, satellite detection seems unlikely.

\citet{Fortney2016} outline several likely possibilities for the 0.1--100~$\mu m$ spectrum of a classical ice giant at $\sim$ 600~AU. For the pressures expected from outgassing, the gross details of the spectrum from an icy super-Earth are probably similar. If the atmosphere is thin, surface reflectance may yield absorption features of various ices.  For a thicker atmosphere, most of the gaseous methane freezes out. The spectrum of an ice giant is then dominated by Rayleigh scattering at short wavelengths, with possible features from clouds, pressure induced H$_2$-H$_2$ absorption, and methane absorption at longer wavelengths. If these features exist and yield an accurate log~$g$, then it may be possible to derive the mean density from log~$g$ and a radius inferred from the broadband spectrum.

Estimating the amount of methane and other volatiles in the outgassed atmosphere of an icy super-Earth requires an accurate assessment of outgassing and freeze-out. Although quantifying these requires a molecular-scale and macro-scale analysis of the inner workings of the icy planet, it may be possible to place some constraints on the abundance of atmospheric CH$_4$. If any icy super-Earth with no atmosphere has a surface temperature of about $20-30$\,K, only H$_2$ is volatile.  If the outflux of H$_2$ is supercritical (see section $4$) a more substantial H$_2$ atmosphere builds with time. The surface temperature increases. When the surface temperature is roughly $50$\,K, CH$_4$ adsorbed onto the surface water ice turns volatile \citep[see adsorption time scales in][]{Levi2011}. Deposits of solid CH$_4$ are also volatile. The extent of these surface CH$_4$ reservoirs is not known and requires further research. As surface temperatures continue to increase, the kinetics of sI clathrate hydrate of CH$_4$ accelerates. Therefore, the dissociation of the latter phase becomes an important mechanism potentially determining the abundance of atmospheric CH$_4$.

\begin{figure}[ht]
\centering
\includegraphics[trim=0.15cm 4cm 0.2cm 5cm , scale=0.60, clip]{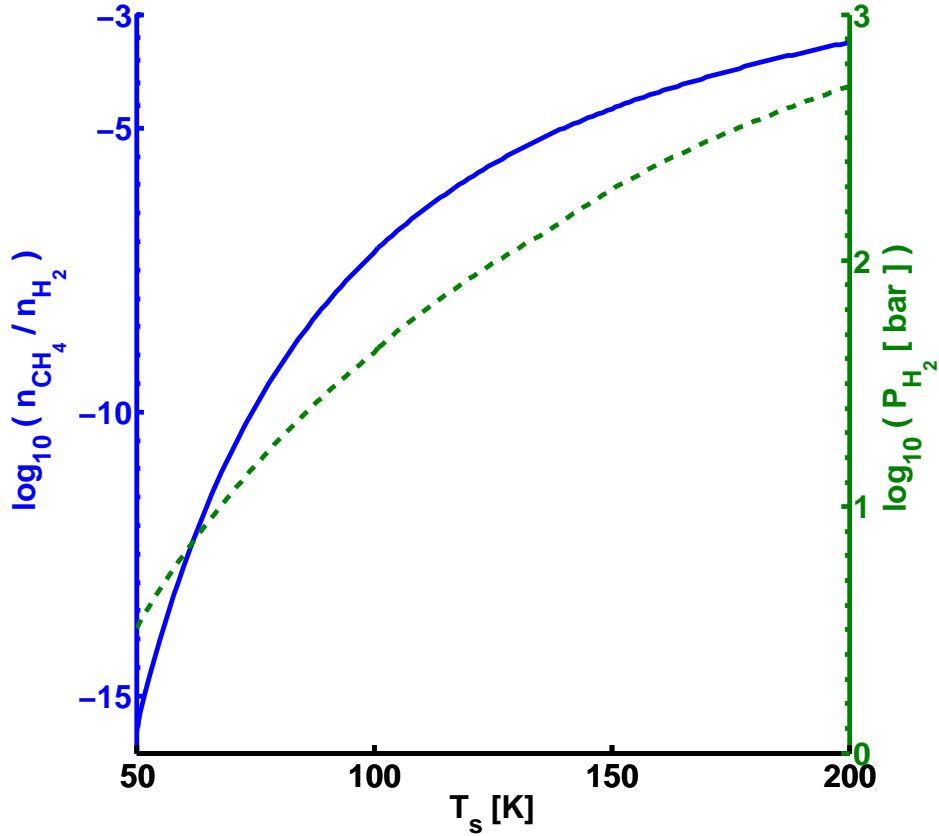}
\caption{\footnotesize{CH$_4$ to H$_2$ mole ratio in the atmosphere (solid blue curve), and the atmospheric pressure of H$_2$ (dashed green curve) versus the surface temperature. Unit opacity is taken to be at $0.1$\,bar and $20$\,K. For higher unit opacity pressures or lower effective temperatures (i.e. temperature at unit opacity) the mole ratio given here is an upper bound value.}}
\label{fig:H2CH4Ratio}
\end{figure}

If sI clathrate hydrates of CH$_4$ become available to the surface, we can derive an approximate atmospheric abundance of CH$_4$ as long as it is a trace constituent of the atmosphere. We set the planetary surface temperature from a pure H$_2$ atmosphere and assume the CH$_4$ content independently attains its clathrate hydrate dissociation pressure. In Fig.\ref{fig:H2CH4Ratio} we present the atmospheric CH$_4$/H$_2$ mole ratio and the atmospheric pressure of H$_2$ as a function of the planetary surface temperature. We assume unit opacity is at $0.1$\,bar and an effective temperature of $20$\,K. For lower effective temperatures or higher unit opacity pressures, more H$_2$ is required in the atmosphere to maintain the same surface temperature. Therefore, the mole ratio of Fig.\ref{fig:H2CH4Ratio} is an upper bound value for these cases.  

To test the assumption that atmospheric CH$_4$ does not contribute substantially to the surface temperature, we use the results of \cite{Pavlov2003}. They showed that increasing the abundance of atmospheric CH$_4$ from $1.7$\,ppm to $100$\,ppm increases the surface temperature by $12$\,K. Since the number density of the greenhouse gas in the atmosphere is the relevant parameter we convert the above to $5\times 10^{-15}$\,K\,cm$^3$\,molec$^{-1}$. For a surface temperature of $115$\,K (corresponding to $70$\,bar of H$_2$, for the above unit opacity parameters) the contribution of CH$_4$ to the surface temperature is about $10$\%.
Thus, our estimate of the CH$_4$/H$_2$ mole ratio is good up to $T_s\approx 100$\,K. The additional effect of CH$_4$ as a greenhouse gas implies that less H$_2$ is needed than derived here to reach surface temperatures appropriate for the formation of a subterranean ocean.
  
This analysis suggests that the atmospheric methane abundance might probe the underlying structure of Planet Nine. Although it is
necessary to build much more comprehensive internal structure calculations of (i) ice giants with H-He atmospheres and (ii) icy super-Earths with outgassed atmospheres, it might be possible to relate the atmospheric abundance of CH$_4$ and other volatiles to the structure of the solid-gas boundary and the deeper internal structure of the planet. Aside from applications to Planet Nine, these analyses might be eventually used to probe the structures of icy exoplanets far from their host stars.

\section{SUMMARY}

When icy planets grow at 250--1000~AU from a solar-type star, they reach super-Earth masses too late to accrete H--He gas from a circumstellar disk. However, high pressure in the mantle of these planets converts CH$_4$ into ethane, butane, or diamond. For planets with masses exceeding 5~\mearth\, the hydrogen released during this conversion can reach the surface and produce an atmosphere with a base pressure of several hundred bars.

For simplified models of the internal structure, the conditions at the base of the atmosphere favor clathrate hydrate formation, where the atmospheric hydrogen is locked in the solid water lattice.  When the outflux of H$_2$ is smaller than the critical rate of roughly $10^{10}$\,molec\,cm$^{-2}$\,s$^{-1}$, the outgassed atmosphere is then dynamically stable for a small base pressure $\lesssim$ 1 bar.  Super-critical outflows can establish a substantial hydrogen atmosphere where the base pressure may approach a maximum level of $10^3 - 10^4$ bar.

The atmospheric structure of icy planets with super-critical outflows of H$_2$ depend on the chemical composition 
of the ice crust and the conditions at unit opacity. In icy planets closer to their host stars (effective temperature
$\sim$ 30~K) or with ice crusts richer in clathrate hydrate promoters (lower thermal conductivity) are more likely
to have a subterranean ocean. This ocean probably prevents fractures in the outer crust, restricting outgassing 
of internal hydrogen. The base pressure in this picture is probably a hydrogen atmosphere of a few hundred bars.

In the atmospheres of icy planets at larger distances from their host stars (effective temperature $\sim$ 10~K) or 
with more pure ice crusts, the hydrogen greenhouse effect may produce temperatures and base-pressures larger than the triple point of 
ice Ih-ice III-liquid water. A subterranean ocean is then much less likely. With no restrictions on the applied stress 
on the crust, it is much easier for internal hydrogen to outgas into the atmosphere. The atmospheric pressure may then 
reach a maximum level of $\sim10^4$ bars.

Observations can test these conclusions.  Together with more comprehensive calculations of the internal structure, 
measurements of the gas-phase abundances of CH$_4$ and other H-rich molecules can probe the atmospheric properties
of Planet Nine.  In a more basic test, Planet Nine candidates with outgassed H$_2$ atmospheres should have much larger 
mean density than those with He-rich atmospheres accreted from the protosolar nebula. Several types of data, including 
broadband spectroscopy, and occultations, can constrain the mean density and yield insights into the origin of any
massive planet in the outer solar system.

\section{ACKNOWLEDGEMENTS}

We wish to thank Prof. Dimitar Sasselov for his helpful suggestions. We further wish to thank our anonymous referee for valuable comments.

\bibliographystyle{aasjournal}
\bibliography{amitbib} 

\end{document}